\documentclass[usegraphicx]{mn2e}

\title[A double molecular disc in NGC 6946]
      {A double molecular disc in the triple-barred starburst galaxy NGC
       6946: structure and stability}

\author[A. B. Romeo and K. Fathi]
       {Alessandro B. Romeo$^{1}$\thanks{E-mail: romeo@chalmers.se}
        and Kambiz Fathi$^{2,3}$\\
        $^{1}$Department of Earth and Space Sciences,
              Chalmers University of Technology,
              SE-41296 Gothenburg, Sweden\\
        $^{2}$Department of Astronomy,
              Stockholm University,
              AlbaNova Centre,
              SE-10691 Stockholm, Sweden\\
        $^{3}$Oskar Klein Centre for Cosmoparticle Physics,
              Stockholm University,
              SE-10691 Stockholm, Sweden}

\begin{document}

\date{Accepted 2015 May 28.
      Received 2015 May 22; in original form 2015 March 04}

\pagerange{\pageref{firstpage}--\pageref{lastpage}}

\pubyear{2015}

\maketitle

\label{firstpage}

\begin{abstract}
The late-type spiral galaxy NGC 6946 is a prime example of molecular gas
dynamics driven by `bars within bars'.  Here we use data from the BIMA SONG
and HERACLES surveys to analyse the structure and stability of its molecular
disc.  Our radial profiles exhibit a clear transition at distance
$R\sim1\,\mbox{kpc}$ from the galaxy centre.  In particular, the surface
density profile breaks at $R\approx0.8\,\mbox{kpc}$ and is well fitted by a
double exponential distribution with scale lengths
$R_{1}\approx200\,\mbox{pc}$ and $R_{2}\approx3\,\mbox{kpc}$, while the 1D
velocity dispersion $\sigma$ decreases steeply in the central kpc and is
approximately constant at larger radii.  The fact that we derive and use the
full radial profile of $\sigma$ rather than a constant value is perhaps the
most novel feature of our stability analysis.  We show that the profile of
the $Q$ stability parameter traced by CO emission is remarkably flat and well
above unity, while the characteristic instability wavelength exhibits clear
signatures of the nuclear starburst and inner bar within bar.  We also show
that CO-dark molecular gas, stars and other factors can play a significant
role in the stability scenario of NGC 6946.  Our results provide strong
evidence that gravitational instability, radial inflow and disc heating have
driven the formation of the inner structures and the dynamics of molecular
gas in the central kpc.
\end{abstract}


\section{INTRODUCTION}

Non-axisymmetric galactic structures such as bars and spiral arms contain a
significant fraction of the disc material, and thus have a prominent role in
the evolution of their host galaxies (e.g., Lindblad 1960).  Such structures
are efficient drivers for the redistribution of angular momentum, and induce
strong flows and streaming motions in their hosts (e.g., Schwarz 1984).  The
interstellar gas is highly inhomogeneous and prone to gravitational
instability, thermal instability and turbulent compression (see, e.g., Mac
Low \& Klessen 2004).  When large-scale flows converge, molecular clouds can
form rapidly in regions dense enough to form $\mathrm{H}_{2}$ (e.g., Glover
\& Mac Low 2007; Tasker 2011).  Galactic-scale flows also pump turbulent
energy into the rotating gas disc, which in turn initiates local
instabilities (e.g., Renaud et al.\ 2013) and produces further heating of the
disc (e.g., Martig et al.\ 2014).  A thorough understanding of such
multi-scale processes requires a detailed analysis of the structures that
originate from them.  Galaxies hosting a hierarchy of interlinked
non-axisymmetric structures are thus especially suitable for analysing disc
instability and its effects on structure formation and evolution.

The late-type spiral galaxy NGC 6946 is a prime example of such dynamics: it
hosts an outer bar that drives the evolution of structure in the disc, from
its prominent spiral arms to an inner bar within bar feeding a nuclear
starburst (e.g., Elmegreen et al.\ 1992; Schinnerer et al.\ 2006; Fathi et
al.\ 2007; Tsai et al.\ 2013).  The gravitational instability of NGC 6946 has
been investigated in several works, most often in the context of large galaxy
surveys (e.g., Kennicutt 1989; Ferguson et al.\ 1998; Fuchs \& von Linden
1998; Martin \& Kennicutt 2001; Leroy et al.\ 2008; Romeo \& Wiegert 2011;
Hoffmann \& Romeo 2012; Romeo \& Falstad 2013).  Ferguson et al.\ (1998)
showed how sensitive the results are to the gas 1D velocity dispersion
$\sigma_{\mathrm{g}}$: assuming that
$\sigma_{\mathrm{g}}=6\;\mbox{km\,s}^{-1}$, the value proposed by Kennicutt
(1989), NGC 6946 turns out to be unstable up to the edge of the optical disc,
while using a radial profile of $\sigma_{\mathrm{g}}$ derived from
observations yields stability across the entire disc!  Martin \& Kennicutt
(2001) pointed out that radial variation in $\sigma_{\mathrm{g}}$ remains
controversial because such measurements demand both high angular resolution
and high brightness sensitivity, requirements not met by most observations.
Fortunately, recent CO and H\,\textsc{i} galaxy surveys (BIMA SONG, HERACLES
and THINGS) have provided high-quality measurements of gas kinematics, which
allow deriving reliable radial profiles of $\sigma_{\mathrm{g}}$ (e.g.,
Cald\'{u}-Primo et al.\ 2013).

Another source of concern in this stability context is the multi-component
(gas+stars) nature of the disc.  Unfortunately, there are no stellar spectra
available for NGC 6946, i.e.\ we have no information about the stellar radial
velocity dispersion $\sigma_{\star}$.  Romeo \& Falstad (2013) used a
model-based radial profile of $\sigma_{\star}$, originally proposed by Leroy
et al.\ (2008), and analysed the stability of NGC 6946 together with other
spirals from The H\,\textsc{i} Nearby Galaxy Survey (THINGS).  They showed
that NGC 6946 is unstable only within the central kpc, and that molecular gas
($\mathrm{H}_{2}$) is the main driver of instability if
$\sigma_{\mathrm{H2}}$ is within the range of values used in previous
investigations, i.e.\ 6 km\,s$^{-1}$ (e.g., Kennicutt 1989; Wilson et
al.\ 2011) to 11 km\,s$^{-1}$ (e.g, Leroy et al.\ 2008).  They also found
that if $\sigma_{\mathrm{H2}}\approx6\;\mbox{km\,s}^{-1}$, then molecular gas
controls the local stability level of the disc up to about 5 kpc from the
centre.  Stars dominate the value of the $Q$ stability parameter at larger
distances, while atomic gas has a negligible effect up to the optical radius.

The multi-component stability analysis of Romeo \& Falstad (2013) has one
main limitation: it uses observationally motivated values of
$\sigma_{\mathrm{H2}}$, rather than a radial profile
$\sigma_{\mathrm{H2}}(R)$ derived from observations.  This is a serious
limitation because, as discussed above, molecular gas plays a primary role
even beyond the central kpc, and up to a radius $R_{\mathrm{max}}$ that
depends on $\sigma_{\mathrm{H2}}$: $R_{\mathrm{max}}\approx5\,\mbox{kpc}$ for
$\sigma_{\mathrm{H2}}\approx6\;\mbox{km\,s}^{-1}$; the larger
$\sigma_{\mathrm{H2}}$, the smaller $R_{\mathrm{max}}$.  This illustrates how
important it would be to have a reliable radial profile of
$\sigma_{\mathrm{H2}}$ up to $R\approx5\,\mbox{kpc}$, and suggests that a
one-component stability analysis based on such a profile could be a proper
starting point for discussing the roles that stars, the various gas phases,
the various bar structures and other components play in the stability
scenario of NGC 6946.

In this paper, we analyse NGC 6946 focusing on the structure and stability of
its molecular disc.  For this purpose, we derive high-quality radial profiles
of the surface density, 1D velocity dispersion and epicyclic frequency up to
$R=5\,\mbox{kpc}$, using data from two recent surveys: the BIMA Survey of
Nearby Galaxies (BIMA SONG, Helfer et al.\ 2003), and the HERA CO-Line
Extragalactic Survey (HERACLES, Leroy et al.\ 2009).  We also analyse the
link between structure formation and gravitational instability in the central
kpc, and discuss the roles that CO-dark molecular gas, stars and other
factors play in the stability scenario of NGC 6946.

\section{NGC 6946}

NGC 6946 is a low-inclination grand-design barred spiral galaxy with an
observed axis ratio of 0.87 (Carignan et al.\ 1990; Zimmer et al.\ 2004;
Boomsma 2007).  We adopt a distance of 5.9 Mpc (Leroy et al.\ 2009), so that
1\arcsec corresponds to 28.6 pc.  Numerous morphological and kinematic
studies by e.g.\ Elmegreen et al.\ (1992), Regan \& Vogel (1995), Elmegreen
et al.\ (1998), Kennicutt et al.\ (2003), Schinnerer et al.\ (2006) and Fathi
et al.\ (2007) have revealed clear signatures of three main gravitational
distortions: a large but weak bar-like structure with a projected ellipticity
of 0.25 and radius $R\approx4.5\arcmin$, a secondary bar with radius
$R\approx60\arcsec$, and a nuclear bar with ellipticity 0.4 and radius
$R\approx8\arcsec$.  Fathi et al.\ (2007, 2009) found that the pattern speed
of the weak outermost bar is
$\Omega_{\mathrm{p}}=25\pm6\;\mbox{km\,s}^{-1}\,\mbox{kpc}^{-1}$, and used
this value to place the corotation radius between 250\arcsec and 300\arcsec,
the outer inner Lindblad resonance (oILR) radius between 30\arcsec and
60\arcsec, and the inner inner Lindblad resonance (iILR) radius below
15\arcsec.  In addition, direct measurements of the pattern speed of the
secondary bar showed that this lies in the range
$39\pm8\;\mbox{km\,s}^{-1}\,\mbox{kpc}^{-1}$ (Zimmer et al.\ 2004),
$47^{+3}_{-2}\;\mbox{km\,s}^{-1}\,\mbox{kpc}^{-1}$ (Fathi et al.\ 2007), and
$51\pm3\;\mbox{km\,s}^{-1}\,\mbox{kpc}^{-1}$ (Font et al.\ 2014).

\begin{figure}
\includegraphics[scale=.84]{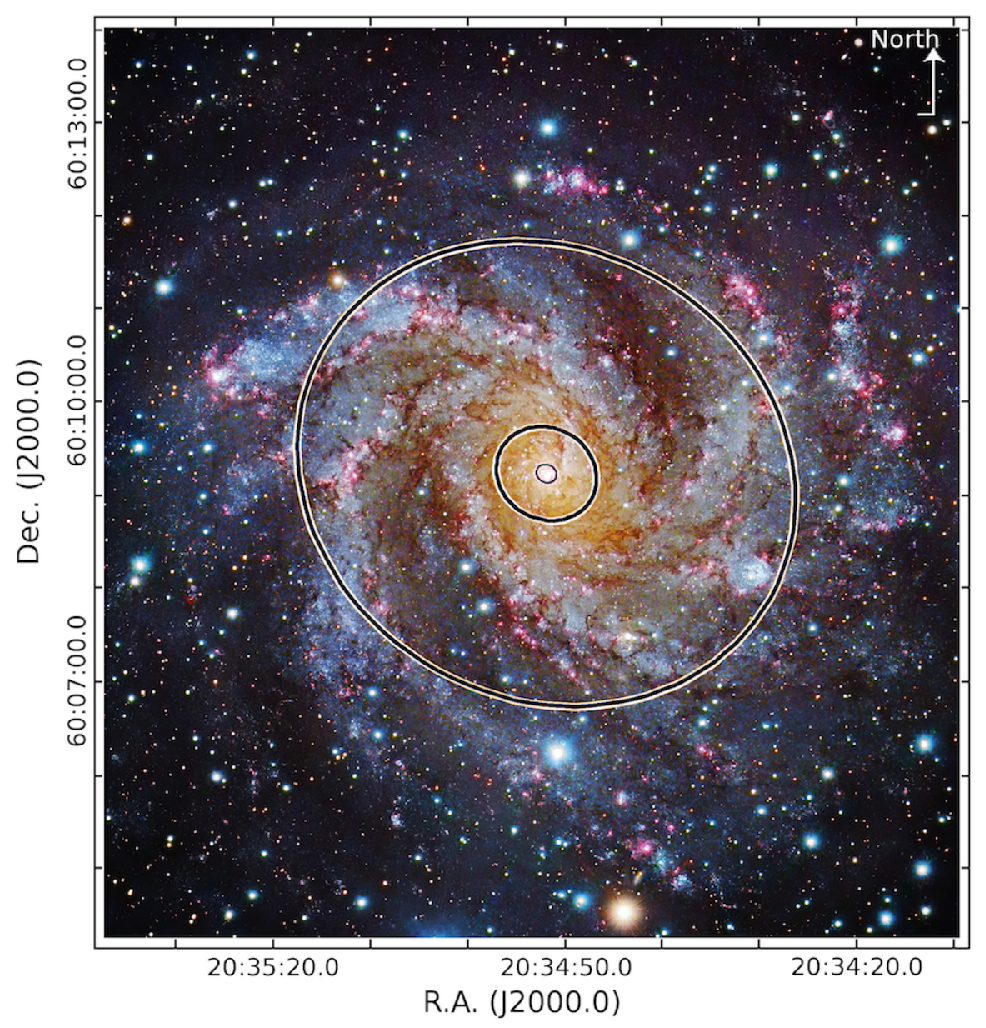}
\caption{A composite $BVI$ image of NGC 6946, with the North-East direction
  displayed in the top-right corner.  The three ellipses show the outermost
  radius of the analysis presented here (5 kpc), and the approximate radii of
  the inner bar within bar (1 kpc and 200 pc).  The ellipses take into
  account the inclination and position angle of the galaxy.  Note how the
  dust lane near the centre in the North direction changes angle at the 1 kpc
  radius, i.e.\ approximately where the oILR of the outer bar is located.
  Image courtesy of Subaru Telescope (NAOJ) and Robert Gendler.}
\end{figure}

These dynamical and structural properties agree well with those found by
Schinnerer et al.\ (2006, 2007), and converge towards a scenario where the
outer bar drives the evolution of structure in NGC 6946 (see Fig.\ 1): the
secondary bar is within the oILR of the outer bar, while the nuclear bar is
within the iILR of the outer bar and has approximately the same radial extent
as the nuclear starburst ($R\approx5.5\arcsec$; Schinnerer et al.\ 2006).
Such structural interplay is indeed interesting from a dynamical point of
view, and makes NGC 6946 an ideal laboratory for a detailed analysis of its
disc structure and stability.

\section{DATA AND METHOD}

\subsection{Molecular gas data}

In this work we analyse the molecular gas content and kinematics of NGC 6946
as traced by the CO $J(1\to0)$ and CO $J(2\to1)$ lines, and based on archival
single-dish and interferometric observations from the BIMA SONG (Helfer et
al.\ 2003) and HERACLES (Leroy et al.\ 2009) surveys, respectively.  We have
retrieved the reduced data cubes from the archives made available by these
teams.  All the details concerning data acquisition, data reduction and data
quality are presented in the two papers above (see also Leroy et al.\ 2008
for the HERACLES data).  Most important for our analysis is that the spatial
resolution of the CO $J(1\to0)$ data is approximately
$5\arcsec\times6\arcsec$, sampled at 1\arcsec/pix and with a channel width of
10 km\,s$^{-1}$.  The CO $J(2\to1)$ data have a spatial resolution of
approximately 11\arcsec, sampled at 2\arcsec/pix and with a channel width of
5.2 km\,s$^{-1}$.

To derive the molecular gas surface density ($\Sigma$), line-of-sight
velocity ($V_{\mathrm{los}}$) and velocity dispersion ($\sigma$), we use two
methods in parallel, namely calculating the moment maps and applying Gaussian
fits to the individual spectra.  We find a good agreement between the moment
maps and the results of Gaussian fitting.  We also find that our amplitude
maps match the zeroth-moment maps presented by the BIMA SONG and HERACLES
teams.  One advantage of applying Gaussian fits is that we can get better
covering maps if we use a Hanning smoothing algorithm in the spectral
direction before fitting the individual Gaussian profiles (e.g., Hernandez et
al.\ 2005; Daigle et al.\ 2006).  This procedure does not affect the line
amplitude or shift.  However, it artificially broadens the individual spectra
by $\Delta\sigma\approx4.4\;\mbox{km\,s}^{-1}$, which we then subtract
quadratically from the derived velocity dispersion maps.  Finally, we apply a
cleaning procedure by removing all the spectra for which the emission line
amplitude is smaller than twice the rms noise, $\sigma$ is smaller than half
the velocity channel, and the formal error in the derived $\sigma$ is greater
than 10 km\,s$^{-1}$.

A careful inspection of the individual spectra reveals the presence of
multiple components at different locations (mostly in the central few 100
pc).  In view of the good agreement between the moment maps and the
single-Gaussian fits, and in view of the complications involved in
multiple-profile fitting schemes (Blasco-Herrera et al.\ 2010), we do not use
profile decomposition methods.  All subsequent analysis is thus based on
single-Gaussian profile fitting.

\subsection{Radial profiles by means of robust statistics}

We derive the radial profiles of $\Sigma$ and $\sigma$ using robust
statistics, which are especially useful when the data contain a significant
fraction of outliers (see, e.g., Rousseeuw 1991; M\"{u}ller 2000; Huber \&
Ronchetti 2009; Feigelson \& Babu 2012).  This is indeed the case for the
$\Sigma$ and $\sigma$ maps, where a non-negligible number of pixel values
(e.g.\ those associated with H\,\textsc{ii} regions) deviate strongly from a
normal distribution.  To derive the radial profiles of $\Sigma$ and $\sigma$,
we divide their maps into tilted rings, which are circular in the plane of
the galaxy, and compute the median values of $\Sigma$ and $\sigma$ in each
ring.  We then estimate the uncertainty in these median values via the median
absolute deviation (MAD):
\begin{equation}
\Delta X_{\mathrm{med}}=1.858\times\mbox{MAD}/\sqrt{N}\,,
\end{equation}
\begin{equation}
\mbox{MAD}=\mbox{median}\{|X_{i}-X_{\mathrm{med}}|\}\,,
\end{equation}
where $X_{i}$ are the individual measurements of $\Sigma$ or $\sigma$,
$X_{\mathrm{med}}$ is their median value, and $N$ is the number of resolution
elements in each ring (i.e.\ the number of pixels in the ring divided by the
number of pixels per resolution element).  Eqs (1) and (2) are the robust
counterparts of the formula traditionally used for estimating the uncertainty
in the mean: $\Delta X_{\mathrm{mean}}=\mbox{SD}/\sqrt{N}$, where SD denotes
the standard deviation (see again M\"{u}ller 2000).  Indeed, the median and
the median absolute deviation provide robust statistical estimates of the
`central value' and the `width' of a data set, respectively, even when almost
50\% of the data are outliers!

\begin{figure}
\includegraphics[scale=.93]{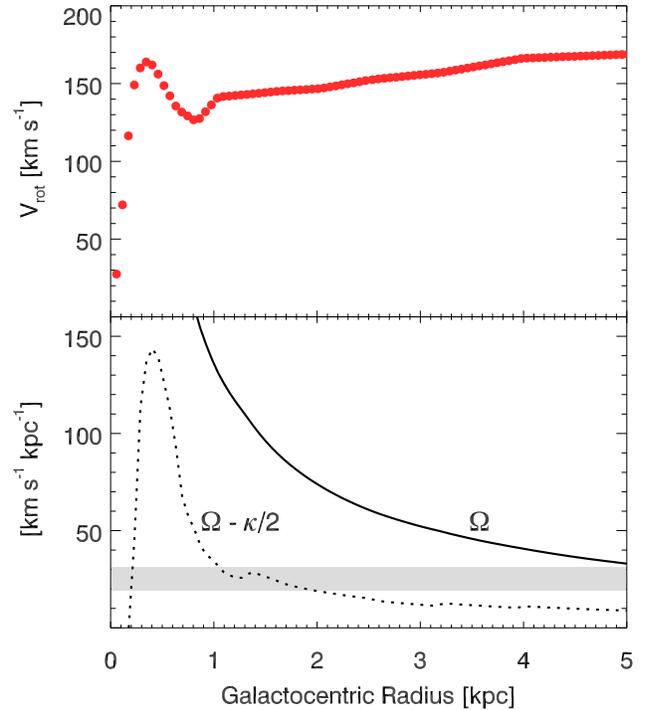}
\caption{The rotation curve (top) and frequency diagram (bottom) of NGC 6946.
  Adopting a pattern speed of $25\pm6\;\mbox{km\,s}^{-1}\,\mbox{kpc}^{-1}$
  for the weak outer bar (Fathi et al.\ 2009), we recover resonance radii
  consistent with the literature.  The rotation curve is consistent with
  measurements based on different molecules, neutral atomic and ionized gas.
  The frequency diagram is based on smoothed curves so as to mimic the
  resolution of the curves presented in the literature (see Sect.\ 2).}
\end{figure}

Our numerous tests with varying ring radii and widths confirm that the high
quality of the BIMA SONG and HERACLES data allow a derivation of $\Sigma$ and
$\sigma$ at radii smaller than the synthesized beam size of the
interferometric observations.  This is mainly thanks to the good sampling of
the resolution element.  The smallest reliable step is found to be 2\arcsec,
which corresponds to 57 pc.  This is the step size adopted throughout our
analysis.  The inner 2.5 kpc of the $\Sigma$ and $\sigma$ profiles presented
here are derived from the BIMA SONG data, while the profiles at larger radii
are derived from the HERACLES data.  The $\mathrm{H}_{2}$ surface density is
converted to physical units by adopting an $X_{\mathrm{CO}}$ conversion
factor of $2\times10^{20}\;\mbox{cm}^{-2}\;(\mbox{K\,km\,s}^{-1})^{-1}$,
consistent with Leroy et al.\ (2008) and Donovan Meyer et al.\ (2012).  We
also correct for the contribution of Helium multiplying by a factor of 1.36.
To further match the BIMA SONG and HERACLES data, we use
$\mbox{CO}\;J(2\to1)/\mbox{CO}\;J(1\to0)=0.8$, as in Leroy et al.\ (2008).

In addition to the $\Sigma$ and $\sigma$ profiles, a stability analysis also
requires the epicyclic frequency $\kappa$ at each radius, which we derive
from the observed rotation curve.  To calculate the rotation curve, we assume
that circular rotation is the dominant kinematic feature, and that our
measurements refer to positions on a single inclined disc.  We then use the
tilted ring method combined with the harmonic decomposition formalism (e.g.,
Schoenmakers et al.\ 1997; Wong et al.\ 2004; Fathi et al.\ 2005).  Given
that this procedure involves fitting several parameters at each radius of the
observed velocity field (contrary to simply finding the median values of
$\Sigma$ and $\sigma$), we cannot use the same initial step size.  Hence we
apply a larger radial step, and interpolate linearly to obtain the rotation
curve at all radii where we have calculated the robust $\Sigma$ and $\sigma$
values.  Here the cut in the data is made at a radius of 0.7 kpc.  Once the
rotation curve $V_{\mathrm{rot}}$ is calculated at each radius $R$, we derive
the angular frequency $\Omega = V_{\mathrm{rot}}/R$ and the epicyclic
frequency $\kappa=\sqrt{R\,d\Omega^{2}/dR+4\Omega^{2}}$.  Setting up the
standard frequency diagram (see Fig.\ 2), we confirm the location of the
inner Lindblad resonances of the weak outer bar discussed in Sect.\ 2.

\section{RESULTS}

\subsection{Structure of the disc}

\begin{figure}
\includegraphics[angle=-90.,scale=.94]{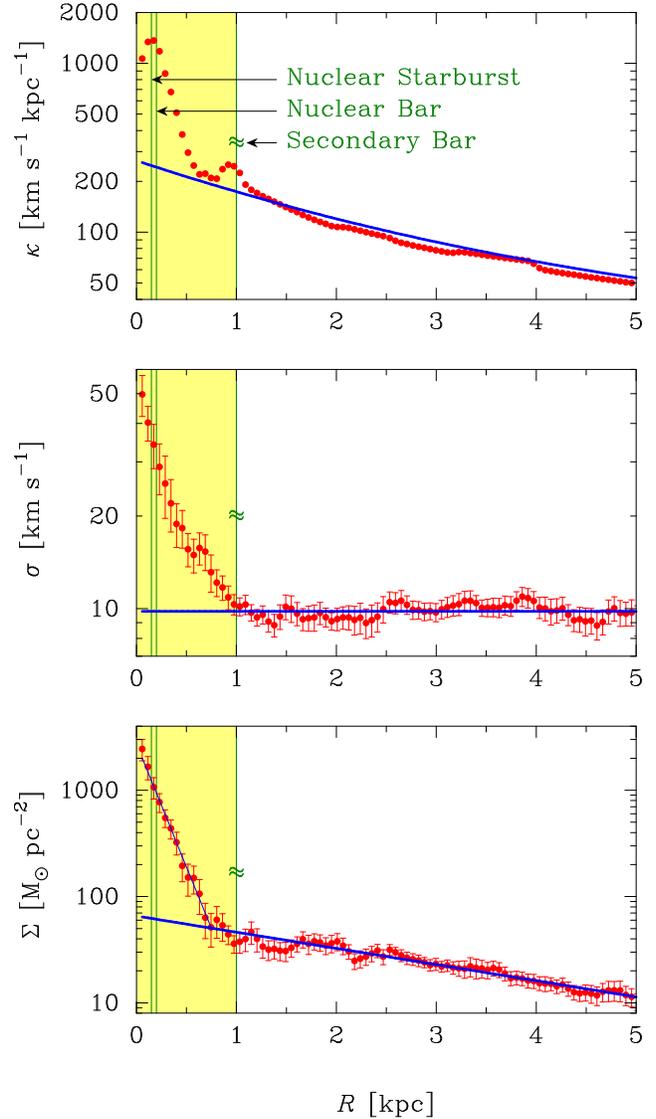}
\caption{Radial profiles of the epicyclic frequency (top), 1D velocity
  dispersion (middle) and surface density (bottom) of molecular gas in NGC
  6946.  The thick lines are fits to these profiles over the radial range
  1--5 kpc, extrapolated into the central kpc.  The thin lines represent the
  approximate radial extent of the nuclear starburst and inner bar within
  bar.  The bottom panel also shows a robust, median-based, exponential fit
  to $\Sigma(R)$ for $R\leq0.8\,\mbox{kpc}$.}
\end{figure}

Fig.\ 3 illustrates that the radial profiles of the epicyclic frequency,
$\kappa$, 1D velocity dispersion, $\sigma$, and surface density, $\Sigma$,
break at $R\sim1\,\mbox{kpc}$.  This tells us that the inner region of NGC
6946 is physically distinct from the rest of the disc, and that radial inflow
(increase of $\Sigma$ and $\kappa$ towards the centre) and disc heating
(increase of $\sigma$ towards the centre) have both played a primary role in
shaping the central kpc.  This is consistent with the fact that the dynamics
of NGC 6946 is driven by three bars, and that the strong secondary bar
extends up to $R\sim1\,\mbox{kpc}$ (see Sect.\ 2).  Since bars are
non-axisymmetric gravitational instabilities, they exert torques on the disc,
which transport angular momentum and energy outwards, leading to the two
processes mentioned above: radial inflow and disc heating%
\footnote{Disc heating is a natural consequence of radial inflow and is
  mediated by local gravitational instabilities.  Although there are still
  open questions, the basic idea behind this process is simple, and is
  beautifully illustrated in sect.\ 7.1 of Kormendy \& Kennicutt (2004).
  Radial inflow increases both $\Sigma$ and $\kappa$, but $\Sigma$ `wins' and
  the Toomre (1964) parameter $Q=\kappa\sigma/\pi G\Sigma$ decreases.  As $Q$
  drops below a critical value of order unity, local gravitational
  instabilities set in and increase $\sigma$, thus heating the disc.}
(e.g., Zhang 1998; Griv et al.\ 2002; Romeo et al.\ 2003, 2004; Fathi et
al.\ 2008; Agertz et al.\ 2009; Krumholz \& Burkert 2010; Forbes et
al.\ 2012, 2014).  Besides, bars within bars make such processes highly
efficient (e.g., Shlosman et al.\ 1989; Friedli \& Martinet 1993; Rautiainen
et al.\ 2002; Shlosman 2002).  A dynamically cool inner disc and especially
nuclear star formation make such structures long-lived, even in the absence
of mode coupling (Du et al.\ 2015; Wozniak 2015).

Look now at the thick lines in Fig.\ 3.  These are simple but accurate fits
to the radial profiles of $\kappa$, $\sigma$ and $\Sigma$ for
$1\,\mbox{kpc}\leq R\leq5\,\mbox{kpc}$.
\begin{enumerate}
\item $\kappa_{\mathrm{fit}}(R)$ is the fit originally proposed by Leroy et
  al.\ (2008); see their eqs (13) and (B1), and table 4.
\item $\sigma_{\mathrm{fit}}(R)\simeq10\;\mbox{km\,s}^{-1}$ is the median of
  the 1D velocity dispersion data.  This value is comparable to those found
  by Walsh et al.\ (2002) and Cald\'{u}-Primo et al.\ (2013) for
  $R\ga2\,\mbox{kpc}$.
\item $\Sigma_{\mathrm{fit}}(R)=\Sigma_{2}\,\mathrm{e}^{-R/R_{2}}$, with
  $\Sigma_{2}\simeq66\;\mbox{M}_{\odot}\,\mbox{pc}^{-2}$ and
  $R_{2}\simeq2.9\,\mbox{kpc}$, is a median-based fit to the surface density
  data.  The disc scale length $R_{2}$ is 50\% larger than that found by
  Leroy et al.\ (2008, 2009).  Note, however, that they fitted an exponential
  function to the whole molecular disc, i.e.\ they included the central kpc,
  which we have shown to be physically distinct from the rest of the disc.
\end{enumerate}
Extrapolating these fits back to $R<1\,\mbox{kpc}$ allows us to predict how
the inner region of NGC 6946 differs dynamically from the rest of the disc.
In particular, we have compelling evidence that the nuclear disc is also
exponential.  A robust, median-based, fit to the radial profile of $\Sigma$
for $R\leq0.8\,\mbox{kpc}$ yields:
$\Sigma_{\mathrm{FIT}}(R)=\Sigma_{1}\,\mathrm{e}^{-R/R_{1}}$, with
$\Sigma_{1}\simeq2700\;\mbox{M}_{\odot}\,\mbox{pc}^{-2}$ and
$R_{1}\simeq190\,\mbox{pc}$.  This is also the approximate radial extent of
the nuclear bar (see Sect.\ 2).  The precise radius at which the surface
density breaks is $R_{\mathrm{break}}\simeq0.75\,\mbox{kpc}$, and the
corresponding density is
$\Sigma_{\mathrm{break}}\simeq50\;\mbox{M}_{\odot}\,\mbox{pc}^{-2}$.  This
means that the nuclear molecular disc extends over four exponential scale
lengths, thus actually more than the main molecular disc!

Our results concerning the nuclear molecular disc of NGC 6946 should be
compared with those of Regan et al.\ (2001).  They showed that the radial
profile of CO surface brightness decreases steeply for $R\la1.5\,\mbox{kpc}$,
then increases inside a transition layer of radial width $\Delta
R\approx1\,\mbox{kpc}$, and finally decreases again (see their fig.\ 2).
They concluded that the CO emission excess in the central kpc looks like a
bulge, while the decrease at larger radii corresponds to an exponential disc.
This is in contrast to our findings.  We have shown that the CO surface
density excess in the central kpc is indeed a nuclear exponential disc, and
that the transition from the nuclear to the main disc is sharp: the slope of
$\Sigma(R)$ changes abruptly across the radius
$R_{\mathrm{break}}\simeq0.75\,\mbox{kpc}$.  In addition, the fact that the
central 1D velocity dispersion is very high,
$\sigma_{0}\approx50\;\mbox{km\,s}^{-1}$, does not tell us that the nuclear
structure is bulge-like, given that the central surface density is as large
as $\Sigma_{0}\approx2500\;\mbox{M}_{\odot}\,\mbox{pc}^{-2}$.  A real upper
limit on the central disc scale height can be estimated by neglecting the
gravity of the stellar disc, and by assuming that the molecular disc is
self-gravitating and isothermal along the vertical direction:
$h_{0}\ll\sigma_{0}^{2}/\pi G\Sigma_{0}\approx\mbox{70--80}\,\mbox{pc}$
(other assumptions would only modify our estimate by a factor of order unity;
see, e.g., van der Kruit \& Freeman 2011).  Since the nuclear disc has scale
length $R_{1}\simeq190\,\mbox{pc}$ and extends up to
$R_{\mathrm{break}}\simeq0.75\,\mbox{kpc}$, we find that
$h_{0}\ll0.4\,R_{1}\approx0.1\,R_{\mathrm{break}}$.  Thus the central 1D
velocity dispersion is consistent with a nuclear molecular disc of moderate
thickness.

Note that $\sigma_{0}\approx50\;\mbox{km\,s}^{-1}$ is comparable to the CO
values found by Schinnerer et al.\ (2006) and to the central stellar velocity
dispersions measured by Engelbracht et al.\ (1996), Ho et al.\ (2009) and
Kormendy et al.\ (2010).  Note also that $R_{2}\approx3\,\mbox{kpc}$ is
comparable to the scale length of the stellar disc found by Kormendy et
al.\ (2010), and that $R_{1}/R_{2}\approx0.07$ is within the range of
pseudobulge-to-disc scale length ratios observed in spiral galaxies (e.g.,
Courteau et al.\ 1996; MacArthur et al.\ 2003; Fisher \& Drory 2008).  As
noted by John Kormendy (private communication), these results are consistent
with the predictions of the general evolution picture reviewed in Kormendy \&
Kennicutt (2004) and Kormendy (2013).

\subsection{Stability of the disc}

\begin{figure}
\includegraphics[angle=-90.,scale=.95]{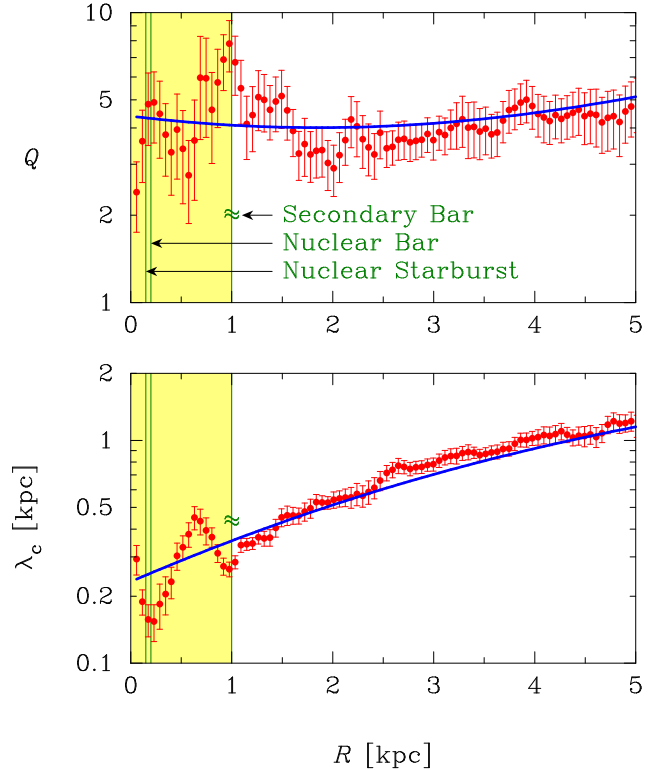}
\caption{Radial profiles of the $\mathcal{Q}$ stability parameter (top) and
  characteristic instability wavelength (bottom) for molecular gas in NGC
  6946.  The thick lines are predictions based on the fits to $\kappa(R)$,
  $\sigma(R)$ and $\Sigma(R)$ shown in Fig.\ 3.  The thin lines represent the
  approximate radial extent of the nuclear starburst and inner bar within
  bar.}
\end{figure}

Fig.\ 4 shows the radial profiles of two useful stability diagnostics.  One
is the Toomre-like parameter
\begin{equation}
\mathcal{Q}=\frac{3}{2}\,\frac{\kappa\sigma}{\pi G\Sigma}\,,
\end{equation}
which measures the stability level of realistically thick gas discs (see
Romeo \& Falstad 2013).%
\footnote{We have assumed that the velocity dispersion is isotropic
  ($\sigma_{z}=\sigma_{R}$).  This is justified by the fact that interstellar
  gas can, to first approximation, be regarded as a collisional component.}
The other is the characteristic instability wavelength,
\begin{equation}
\lambda_{\mathrm{c}}=2\pi\,\frac{\sigma}{\kappa}\,,
\end{equation}
i.e.\ the scale at which the disc becomes locally unstable as $\mathcal{Q}$
drops below unity (see again Romeo \& Falstad 2013).  Fig.\ 4 also shows
$\mathcal{Q}_{\mathrm{fit}}(R)$ and $\lambda_{\mathrm{c\,fit}}(R)$, which are
predictions based on the fits to the radial profiles of $\kappa$, $\sigma$
and $\Sigma$ (thick lines).

The top panel of Fig.\ 4 illustrates that $\mathcal{Q}(R)$ is fairly constant
for $R\ga1\,\mbox{kpc}$, while it varies significantly for
$R\la1\,\mbox{kpc}$.  This is not surprising since the central kpc is the
region where $\kappa(R)$, $\sigma(R)$ and $\Sigma(R)$ vary most.  As we move
towards the centre, $\kappa$ and $\sigma$ increase by a factor of 5--7, while
$\Sigma$ increases by two orders of magnitude (see Fig.\ 3).  What is
surprising is that, in spite of such strong variations, $\mathcal{Q}(R)$
fluctuates around a constant value, deviating from it by less than a factor
of two!  What drives the radial variation of $\mathcal{Q}$ in the central
kpc?  We will clarify this point in Sect.\ 5.1.  Here instead we focus on the
overall flatness of $\mathcal{Q}(R)$, which is an eloquent example of
self-regulation: the disc of NGC 6946 is driven by processes that keep it at
a fairly constant stability level, except in the central kpc, where the value
of $\mathcal{Q}$ is modulated by the nuclear starburst and inner bar within
bar.  Current dynamical models of star-forming galaxies suggest that such
self-regulation processes are determined by a balance among gravitational
instability, star formation and accretion (e.g., Bertin \& Romeo 1988; Romeo
1990; Agertz et al.\ 2009; Krumholz \& Burkert 2010; Elmegreen 2011; Cacciato
et al.\ 2012; Forbes et al.\ 2012, 2014).  According to these models,
galactic discs should (i) have a constant stability level, and (ii) be close
to marginal instability.  The results illustrated above are clearly
consistent with point (i), while the fact that the $\mathcal{Q}$ stability
parameter is well above unity ($\mathcal{Q}_{\mathrm{fit}}=\mbox{4--5}$)
seems to tell us that the disc of NGC 6946 is fairly far from marginal
instability.  Is $\mathcal{Q}$ really so large?  This is a complex issue that
we will discuss in Sect.\ 5.2.

Our results concerning $\mathcal{Q}(R)$ seem comparable to those of Romeo \&
Falstad (2013), who showed that the stability level of THINGS spirals is, on
average, remarkably flat and well above unity (see their figs 4 and 5).
Note, however, that those results cannot be directly compared with ours
because they concern the THINGS sample as a whole, not each of the spirals or
in particular NGC 6946.

The bottom panel of Fig.\ 4 illustrates that $\lambda_{\mathrm{c}}(R)$
increases monotonically for $R\ga1\,\mbox{kpc}$, while it has two
well-defined minima for $R\la1\,\mbox{kpc}$.  The location of such minima is
strongly correlated with the radial extent of the three structures in the
central kpc: the nuclear starburst (NS), the nuclear bar (NB) and the
secondary bar (SB).  We find that:
\begin{equation}
R_{\mathrm{min1}}\approx\mbox{150--200}\,\mbox{pc}\approx
R_{\mathrm{NS}},R_{\mathrm{NB}}\,;
\end{equation}
\begin{equation}
R_{\mathrm{min2}}\approx1\,\mbox{kpc}\approx
R_{\mathrm{SB}}\,.
\end{equation}
This is surprising because $\lambda_{\mathrm{c}}(R)$ is meant to be a
diagnostic for characterizing \emph{local} gravitational instabilities (such
as those associated with star-forming rings, complexes/clumps and
starbursts), and not \emph{global} gravitational instabilities (such as those
classically associated with the formation of spiral and bar structures).
Thus Eqs (5) and (6) point out $\lambda_{\mathrm{c}}(R)$ as a powerful
diagnostic, which accurately predicts the sizes of starbursts and bars within
bars.  In contrast, remember that $\mathcal{Q}(R)$ is only modulated by such
structures.  What drives the radial variation of $\lambda_{\mathrm{c}}$ in
the central kpc?  We will clarify this point in Sect.\ 5.1.  Here instead we
go on with another important result:
\begin{equation}
\lambda_{\mathrm{c}}(R_{\mathrm{min1}})\approx
R_{\mathrm{NS}},R_{\mathrm{NB}}\,;
\end{equation}
\begin{equation}
\lambda_{\mathrm{c}}(R_{\mathrm{min2}})\approx
\frac{1}{4}\,R_{\mathrm{SB}}\,.
\end{equation}
Note that $\lambda_{\mathrm{c}}(R)$ matches the radial extent of both the
nuclear starburst and the nuclear bar, while it is much shorter than the
secondary bar.  This result indicates that the nuclear structures of NGC 6946
are intimately related and were most likely generated by the same process of
local gravitational instability in the disc, while the strong secondary bar
has grown significantly after the onset of instability or was generated by
larger-scale processes.

\section{DISCUSSION}

\subsection{What drives the radial variations of $\mathcal{Q}$ and
            $\lambda_{\mathrm{c}}$ in the central kpc?}

In this section, we clarify which physical processes drive the radial
variations of $\mathcal{Q}$ and $\lambda_{\mathrm{c}}$ in the central kpc of
NGC 6946.  For this purpose, we need to understand in more detail how
$\kappa$, $\sigma$ and $\Sigma$ contribute to such variations.  We can easily
highlight the contribution of, e.g., $\kappa$ by replacing $(\sigma,\Sigma)$
with $(\sigma_{\mathrm{fit}},\Sigma_{\mathrm{fit}})$ in Eqs (3) and (4),
where $\sigma_{\mathrm{fit}}(R)$ and $\Sigma_{\mathrm{fit}}(R)$ are specified
in items (ii) and (iii) of Sect.\ 4.1.  Remember, in fact, that using these
fits for $R<1\,\mbox{kpc}$ suppresses the fundamental differences (in
$\sigma$ and $\Sigma$) between the central kpc and the rest of the disc.  The
contributions of $\sigma$ and $\Sigma$ can be highlighted in a similar way.

\begin{figure}
\includegraphics[angle=-90.,scale=.93]{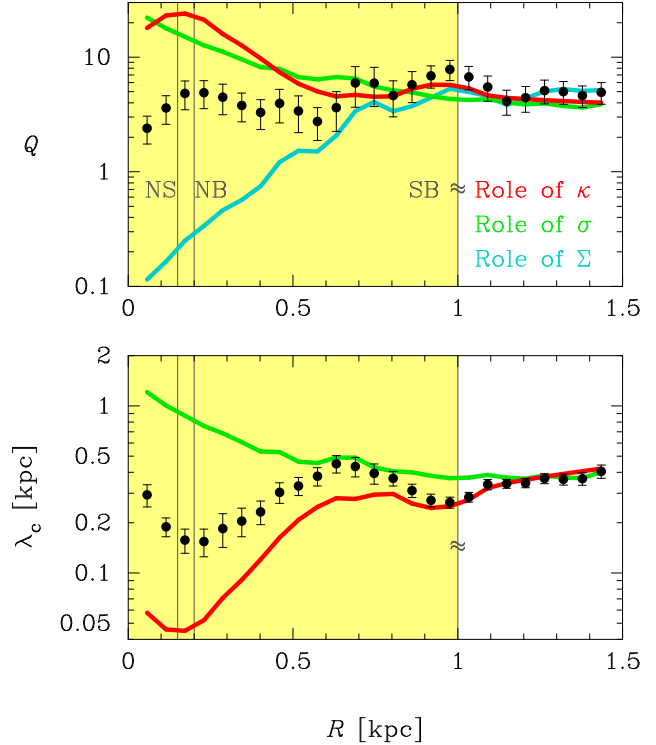}
\caption{Radial profiles of the $\mathcal{Q}$ stability parameter (top) and
  characteristic instability wavelength (bottom) in the inner region of NGC
  6946.  The thick lines show the roles of $\kappa$, $\sigma$ and $\Sigma$ in
  driving the radial variations of $\mathcal{Q}$ and $\lambda_{\mathrm{c}}$.
  The thin lines represent the approximate radial extent of the nuclear
  starburst (NS), nuclear bar (NB) and secondary bar (SB).}
\end{figure}

Fig.\ 5 shows the contributions defined above (thick lines), together with
the original radial profiles of $\mathcal{Q}$ and $\lambda_{\mathrm{c}}$.
The top panel illustrates that $\kappa$ and $\Sigma$ play a more significant
role than $\sigma$ in driving the \emph{radial variation} of $\mathcal{Q}$:
$\kappa$ gives rise to the maximum at $R\approx\mbox{150--200}\,\mbox{pc}$,
while $\Sigma$ contributes to the fluctuations at larger radii.  On the other
hand, $\sigma$ has a significant impact on the \emph{magnitude} of
$\mathcal{Q}$: it helps $\kappa$ to counteract the strongly destabilizing
effect of $\Sigma$, hence to flatten $\mathcal{Q}(R)$.  The bottom panel
illustrates that $\kappa$ shapes the radial profile of $\lambda_{\mathrm{c}}$
more faithfully than that of $\mathcal{Q}$.  However, even in this case,
$\sigma$ has a strong impact on the magnitude of $\lambda_{\mathrm{c}}$.  In
fact, look at the lower curve.  Its minimum at
$R\approx\mbox{150--200}\,\mbox{pc}$ is 4 times deeper than that of
$\lambda_{\mathrm{c}}(R)$.  This means that ignoring the increase of $\sigma$
towards the centre, i.e.\ assuming that $\sigma\approx10\;\mbox{km\,s}^{-1}$
for all $R$, would underpredict the size of the nuclear starburst and bar by
as much as a factor of 4.

The bottom line is that radial inflow (as characterized by $\kappa$ and
$\Sigma$) is the main driver for the radial variations of $\mathcal{Q}$ and
$\lambda_{\mathrm{c}}$ in the central kpc.  Nevertheless, disc heating (as
characterized by $\sigma$) is of fundamental importance for flattening the
stability level of the disc and for regulating the size of the nuclear
structures.

\subsection{Is $\mathcal{Q}$ really so large?}

In this section, we discuss the physical factors that can push the value of
$\mathcal{Q}$ down, closer to unity.  To estimate the magnitude of such
effects and impose tighter constraints on the stability level of NGC 6946, we
need a more general Toomre-like parameter.  Romeo \& Wiegert (2011)
introduced a simple and accurate approximation for the two-component
(stars+gas) $Q$ parameter, which takes into account the stabilizing effect of
disc thickness and predicts whether the local stability level is dominated by
stars or gas.  Romeo \& Falstad (2013) generalized this approximation to
discs made of several stellar and/or gaseous components, and to the whole
range of velocity dispersion anisotropy ($\sigma_{z}/\sigma_{R}$) observed in
galactic discs.  In the two-component case, the $Q$ stability parameter is
given by
\begin{equation}
\frac{1}{\mathcal{Q}}=
\left\{\begin{array}{ll}
       {\displaystyle\frac{W}{T_{1}Q_{1}}+
                     \frac{1}{T_{2}Q_{2}}}
                       & \mbox{if\ \ }T_{1}Q_{1}\geq
                                      T_{2}Q_{2}\,, \\
                       &                            \\
       {\displaystyle\frac{1}{T_{1}Q_{1}}+
                     \frac{W}{T_{2}Q_{2}}}
                       & \mbox{if\ \ }T_{2}Q_{2}\geq
                                      T_{1}Q_{1}\,,
       \end{array}
\right.
\end{equation}
where $Q_{i}=\kappa\sigma_{i}/\pi G\Sigma_{i}$ is the Toomre parameter of
component $i$, $\sigma$ denotes the radial velocity dispersion, and $T_{i}$
and $W$ are given by
\begin{equation}
T_{i}\approx
\left\{\begin{array}{ll}
       {\displaystyle1+0.6\left(\frac{\sigma_{z}}{\sigma_{R}}\right)_{i}^{2}}
                       & \mbox{for\ }0\la(\sigma_{z}/\sigma_{R})_{i}\la0.5\,,
                         \\
                       & \\
       {\displaystyle0.8+0.7\left(\frac{\sigma_{z}}{\sigma_{R}}\right)_{i}}
                       & \mbox{for\ }0.5\la(\sigma_{z}/\sigma_{R})_{i}\la1\,,
       \end{array}
\right.
\end{equation}
\begin{equation}
W=
\frac{2\sigma_{1}\sigma_{2}}
     {\sigma_{1}^{2}+\sigma_{2}^{2}}\,.
\end{equation}
This set of equations tells us that the local stability level of the disc is
dominated by the component with smaller $TQ$.  The contribution of the other
component is suppressed by the $W$ factor.  Note, in particular, that if
$i=1$ is the dominant component and $\sigma_{1}$ differs significantly from
$\sigma_{2}$, then $\mathcal{Q}\approx T_{1}Q_{1}$ (as in the one-component
case).  We are now ready to discuss the physical factors that can push the
disc of NGC 6946 closer to marginal instability.

\begin{figure}
\includegraphics[scale=.93]{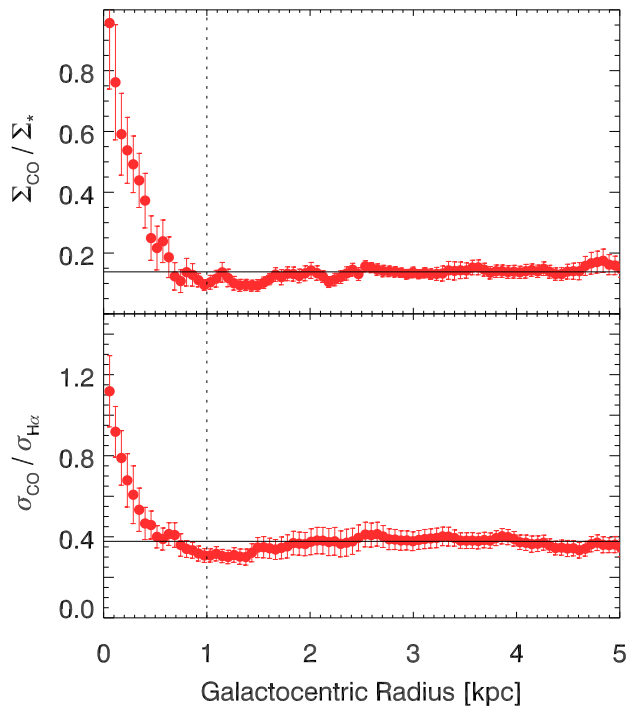}
\caption{Radial profiles of the ratios between the surface densities of
  molecular gas and stars (top) and the 1D velocity dispersions of molecular
  and ionized gas (bottom) in NGC 6946.  The solid lines are the median
  values of $\Sigma_{\mathrm{co}}/\Sigma_{\star}\;(\simeq0.14)$ and
  $\sigma_{\mathrm{co}}/\sigma_{\mathrm{H}\alpha}\;(\simeq0.38)$ for
  $R\leq5\,\mbox{kpc}$.}
\end{figure}

\begin{enumerate}
\item \emph{Stars.}  The condition that $\mathcal{Q}$ is dominated by stars
  ($\star$) rather than molecular gas (co),
  $T_{\star}Q_{\star}<T_{\mathrm{co}}Q_{\mathrm{co}}$, can be written as
  \begin{equation}
  \sigma_{\star}<\sigma_{\mathrm{co}}
                 \left(\frac{\Sigma_{\star}}{\Sigma_{\mathrm{co}}}\right)
                 \left(\frac{T_{\mathrm{co}}}{T_{\star}}\right)\,.
  \end{equation}
  Gerssen \& Shapiro Griffin (2012) showed that
  $(\sigma_{z}/\sigma_{R})_{\star}$ decreases markedly from early- to
  late-type spirals.  For an Sc galaxy like NGC 6946, their best-fitting
  model yieds $(\sigma_{z}/\sigma_{R})_{\star}\approx0.4$ (see their
  fig.\ 4).  Hence $T_{\star}\approx1.1$.  If gas is collisional, as is
  generally assumed, then $(\sigma_{z}/\sigma_{R})_{\mathrm{co}}\approx1$ and
  $T_{\mathrm{co}}\approx1.5$ (see Sect.\ 4.2).  To estimate
  $\Sigma_{\star}/\Sigma_{\mathrm{co}}$, we retrieve the Spitzer IRAC 3.6
  $\mu$m image observed by the SINGS team (Kennicutt et al.\ 2003), and
  derive the radial profile of $\Sigma_{\star}$ using the same method as for
  $\Sigma_{\mathrm{co}}$ (see Sect.\ 3.2) and adopting a fixed $K$-band
  mass-to-light ratio,
  $\Upsilon_{\star}^{K}=0.5\;\mbox{M}_{\odot}/\mbox{L}_{\odot,K}$ (Leroy et
  al.\ 2008).  The top panel of Fig.\ 6 illustrates that
  $\Sigma_{\star}/\Sigma_{\mathrm{co}}\approx7$ for $R\ga1\,\mbox{kpc}$,
  which means that beyond the central kpc the stellar and molecular discs
  have approximately the same scale length.%
  \footnote{Leroy et al.\ (2008) found that the stellar disc scale length is
    30\% larger than the molecular one, but again their exponential fits
    extend over the central kpc, which is physically distinct from the rest
    of the disc.}
  Remembering that $\sigma_{\mathrm{co}}\approx10\;\mbox{km\,s}^{-1}$ for
  $R\ga1\,\mbox{kpc}$ (see Fig.\ 3), the condition that stars dominate over
  molecular gas finally becomes $\sigma_{\star}\la100\;\mbox{km\,s}^{-1}$.
  Since there are no stellar spectra available for NGC 6946, we check this
  condition against the model-based radial profile of $\sigma_{\star}$
  proposed by Leroy et al.\ (2008).  We find that
  $\sigma_{\star}\la100\;\mbox{km\,s}^{-1}$ for $R\ga2\,\mbox{kpc}$,
  therefore stars can actually dominate the value of $\mathcal{Q}$.  But this
  is not the end of the story.  To push the value of $\mathcal{Q}$ down, from
  $T_{\mathrm{co}}Q_{\mathrm{co}}\approx\mbox{4--5}$ to
  $T_{\star}Q_{\star}\approx1$, we should have
  $\sigma_{\star}\approx20\;\mbox{km\,s}^{-1}$, which is true only at the
  edge of the optical disc: $R\approx11\,\mbox{kpc}$.  Even a value of
  $\mathcal{Q}\approx2$ would require
  $\sigma_{\star}\approx\mbox{40--50}\;\mbox{km\,s}^{-1}$ and occur at
  $R\approx7\,\mbox{kpc}$, i.e.\ still beyond the radial range of our
  analysis.  Thus stars alone cannot solve the problem, unless the model
  proposed by Leroy et al.\ (2008) overestimates the real $\sigma_{\star}(R)$
  in NGC 6946.  This underlines the importance of having observed, rather
  than model-based, radial profiles of $\sigma_{\star}$ when analysing the
  stability of disc galaxies.
\item \emph{Atomic gas.}  Detailed comparative analyses of atomic and
  molecular gas in NGC 6946 show that
  $\Sigma_{\mathrm{HI}}\ll\Sigma_{\mathrm{co}}$ for $R\leq5\,\mbox{kpc}$ (see
  fig.\ 40 of Leroy et al.\ 2008), while
  $\sigma_{\mathrm{HI}}\approx\sigma_{\mathrm{co}}$ (see fig.\ 4 of
  Cald\'{u}-Primo et al.\ 2013).  As $T_{\mathrm{HI}}\approx
  T_{\mathrm{co}}(\approx1.5)$, this implies that
  $T_{\mathrm{HI}}Q_{\mathrm{HI}}\gg T_{\mathrm{co}}Q_{\mathrm{co}}$ for
  $R\leq5\,\mbox{kpc}$, i.e.\ atomic gas cannot dominate the value of
  $\mathcal{Q}$ within the radial range of our analysis.
\item \emph{CO-dark molecular gas.}  There is growing evidence that a large
  fraction of molecular gas in the Milky Way is dark, i.e.\ not traced by CO
  emission: $M_{\mathrm{dark}}\approx\mbox{0.4--1.6}\,M_{\mathrm{co}}$ (e.g.,
  Paradis et al.\ 2012; Pineda et al.\ 2013; Kamenetzky et al.\ 2014; Langer
  et al.\ 2014; Smith et al.\ 2014 and references therein; Chen et
  al.\ 2015).  Can CO-dark molecular gas play a significant role in the
  stability scenario of NGC 6946?  To answer this question, we need at least
  rough estimates of both $\Sigma_{\mathrm{dark}}$ and
  $\sigma_{\mathrm{dark}}$.  We estimate $\Sigma_{\mathrm{dark}}$ from the
  mass fraction found in the Milky Way:
  $\Sigma_{\mathrm{dark}}\sim\mbox{0.4--1.6}\,\Sigma_{\mathrm{co}}$.  The
  investigations above also suggest that dark molecular gas is warmer and
  more diffuse than the component traced by CO emission.  So
  $\sigma_{\mathrm{dark}}$ must be bound by
  $\sigma_{\mathrm{co}}(\approx\sigma_{\mathrm{HI}})$ and the 1D velocity
  dispersion of ionized gas:
  $\sigma_{\mathrm{co}}\la\sigma_{\mathrm{dark}}\la
  \sigma_{\mathrm{H}\alpha}$.  We derive the radial profile of
  $\sigma_{\mathrm{H}\alpha}$ from the Fabry-Perot observations of NGC 6946
  presented by Fathi et al.\ (2007), following the same method as for
  $\sigma_{\mathrm{co}}$ (see Sect.\ 3.2).  The bottom panel of Fig.\ 6
  illustrates that
  $\sigma_{\mathrm{H}\alpha}\approx2.6\,\sigma_{\mathrm{co}}$ for
  $R\ga1\,\mbox{kpc}$, while $\sigma_{\mathrm{H}\alpha}$ approaches
  $\sigma_{\mathrm{co}}$ towards the centre.%
  \footnote{This result agrees qualitatively with that found by Westfall et
    al.\ (2014), namely that $\sigma_{\mathrm{H}\alpha}$ is on average twice
    as large as $\sigma_{\mathrm{co}}$ in nearby spiral galaxies.}
  Hence $\sigma_{\mathrm{co}}\la\sigma_{\mathrm{dark}}\la2.6\,
  \sigma_{\mathrm{co}}$.  As $T_{\mathrm{dark}}\approx
  T_{\mathrm{co}}(\approx1.5)$, we find that
  $T_{\mathrm{dark}}Q_{\mathrm{dark}}\sim\mbox{0.6--7}\;T_{\mathrm{co}}
  Q_{\mathrm{co}}$.  This means that CO-dark molecular gas can dominate the
  value of $\mathcal{Q}$ and reduce it by more than 40\%, but only if
  $\sigma_{\mathrm{dark}}\sim\sigma_{\mathrm{co}}$ and
  $\Sigma_{\mathrm{dark}}\sim1.6\,\Sigma_{\mathrm{co}}$, the upper bound
  observed in the Milky Way (Paradis et al.\ 2012).
\item \emph{Anisotropy of the gas velocity dispersion.}  As noted in
  Sect.\ 4.2, interstellar gas is generally regarded as a collisional
  component, so that $(\sigma_{z}/\sigma_{R})_{\mathrm{gas}}\approx1$ and
  $T_{\mathrm{gas}}\approx1.5$.  Bournaud et al.\ (2010) analysed the gas
  velocity field in high-resolution simulations of gas-rich galaxies, and
  found that the velocity dispersion is isotropic only at scales smaller than
  the disc scale height.  Agertz et al.\ (2009) showed that at scales of a
  few 100 pc the gas velocity dispersion has a degree of anisotropy similar
  to the stellar velocity dispersion.  If this is true for NGC 6946, then
  $(\sigma_{z}/\sigma_{R})_{\mathrm{gas}}\approx0.4$ and
  $T_{\mathrm{gas}}\approx1.1$ [see point (i) above], which implies a 30\%
  reduction in the value of $\mathcal{Q}$.
\item \emph{Gas turbulence.}  Turbulence plays a fundamental role in the
  dynamics and structure of cold interstellar gas (see, e.g., Agertz et
  al.\ 2009; Hennebelle \& Falgarone 2012; Falceta-Gon\c{c}alves et
  al.\ 2014; Roy 2015).  The most basic aspect of gas turbulence is the
  presence of supersonic motions.  These are usually taken into account by
  identifying $\sigma_{\mathrm{gas}}$ with the typical 1D velocity dispersion
  of the medium, rather than with its thermal sound speed.  Another important
  aspect of gas turbulence is the existence of scaling relations between
  $\Sigma_{\mathrm{gas}}$, $\sigma_{\mathrm{gas}}$ and the size of the region
  over which such quantities are measured ($\ell$).  Observations show that
  $\Sigma_{\mathrm{HI}}\sim\ell^{1/3}$ and
  $\sigma_{\mathrm{HI}}\sim\ell^{1/3}$ up to scales of 1--10 kpc, whereas
  $\Sigma_{\mathrm{H2}}\sim constant$ and
  $\sigma_{\mathrm{H2}}\sim\ell^{1/2}$ up to scales of about 100 pc (see,
  e.g., Hennebelle \& Falgarone 2012; Falceta-Gon\c{c}alves et al.\ 2014;
  Romeo \& Agertz 2014; Roy 2015).  Motivated by the large observational
  uncertainties of $\Sigma_{\mathrm{gas}}(\ell)$ and
  $\sigma_{\mathrm{gas}}(\ell)$, and having in mind near-future applications
  to high-redshift galaxies, Romeo et al.\ (2010) considered more general
  scaling relations, $\Sigma_{\mathrm{gas}}\propto\ell^{a}$ and
  $\sigma_{\mathrm{gas}}\propto\ell^{b}$, and explored the effect of
  turbulence on the gravitational instability of gas discs.  They showed that
  turbulence excites a rich variety of stability regimes, several of which
  have no classical counterpart.  See in particular the `stability map of
  turbulence' (fig.\ 1 of Romeo et al.\ 2010), which illustrates such
  stability regimes and populates them with observations, simulations and
  models of gas turbulence.  Hoffmann \& Romeo (2012) extended this
  investigation to two-component discs of stars and gas, and analysed the
  stability of THINGS spirals.  They showed that gas turbulence alters the
  condition for star-gas decoupling and increases the least stable
  wavelength, but hardly modifies the $Q$ parameter at scales larger than
  about 100 pc.  Such predictions have now been confirmed and strengthened by
  the simulation work of Agertz et al.\ (2015).
\item \emph{Non-axisymmetric perturbations.}  Like the original Toomre
  parameter, $\mathcal{Q}$ measures the stability of the disc against local
  axisymmetric perturbations, so it assumes that $kR\gg1$.  Here the relevant
  $k$ is the characteristic instability wavenumber,
  $k_{\mathrm{c}}=2\pi/\lambda_{\mathrm{c}}$, so the condition $kR\gg1$ can
  be written as $\lambda_{\mathrm{c}}\ll2\pi R$, which is easy to check (see
  Fig.\ 4).  While this short-wavelength approximation is satisfied by most
  spiral galaxies (Romeo \& Falstad 2013), the assumption of axisymmetric (or
  tightly wound) perturbations is not so general.  Local non-axisymmetric
  stability criteria are far more complex than Toomre's criterion: they
  depend critically on how tightly wound the perturbations are, and cannot
  generally be expressed in terms of a single effective $Q$ parameter (e.g.,
  Lau \& Bertin 1978; Morozov \& Khoperskov 1986; Bertin et al.\ 1989; Jog
  1992; Lou \& Fan 1998; Griv \& Gedalin 2012).  However, there is a general
  consensus that non-axisymmetric perturbations have a destabilizing effect,
  i.e.\ a disc with $1\leq Q<Q_{\mathrm{crit}}$ is still locally unstable
  against such perturbations.  Griv \& Gedalin (2012) found that the usual
  estimate $Q_{\mathrm{crit}}\approx2$ (see, e.g., Binney \& Tremaine 2008)
  is a real upper limit on the critical stability level.  This implies that
  non-axisymmetric perturbations reduce the value of $\mathcal{Q}$ by a
  factor $1/Q_{\mathrm{crit}}\ll1/2$, i.e.\ by much less than 50\%.  It is
  highly non-trivial to constrain the magnitude of this effect more tightly
  than so.  The estimates presented in the literature (see references above)
  depend critically not only on how open the spiral perturbations are, but
  also on the mathematical treatment of the problem.  Tighter constraints on
  the relative importance of non-axisymmetric and axisymmetric perturbations
  in spiral galaxies might be found by analysing the radial profile of
  $\mathcal{Q}$ for a large galaxy sample, and by searching for trends in
  $\mathcal{Q}(R)$ along the Hubble sequence.  This is however well beyond
  the scope of the present paper.
\end{enumerate}

The bottom line of this discussion is that stars, CO-dark molecular gas,
anisotropy of the gas velocity dispersion and non-axisymmetric perturbations
can all play a significant role in the stability scenario of NGC 6946.  The
disc of NGC 6946 can indeed be close to marginal instability.  Nevertheless,
this is most likely the result of all such factors together.  None of such
factors alone can solve the problem.

Finally, note that the points discussed in this section complement, \emph{but
  do not invalidate}, the simple and predictive stability analysis carried
out in Sect.\ 4.2.  In particular, the radial profile of $\mathcal{Q}$
remains remarkably flat.  This is because the main effect here is a
systematic reduction in the value of $\mathcal{Q}$ by a factor of order
unity.  The radial variation of $\mathcal{Q}$ is less significantly affected,
so it is still much weaker than that of $\Sigma$, $\kappa$ or $\sigma$.

\section{CONCLUSIONS}

In this paper, we have provided strong evidence that NGC 6946 has a double
(nuclear+main) molecular disc, with a clear transition at
$R\sim1\,\mbox{kpc}$.  We have analysed the disc structure and stability in
detail, using robust statistics as well as reliable and predictive
diagnostics.  Our major conclusions are pointed out below.
\begin{itemize}
\item The nuclear molecular disc is exponential, and has central surface
  density $\Sigma_{1}\simeq2700\;\mbox{M}_{\odot}\,\mbox{pc}^{-2}$ and scale
  length $R_{1}\simeq190\,\mbox{pc}$.  This is also the approximate radial
  extent of the nuclear bar.  The 1D velocity dispersion decreases steeply
  from about 50 km\,s$^{-1}$ at the galaxy centre to about 10 km\,s$^{-1}$ at
  the outskirts of the disc, i.e.\ at a distance of about 1 kpc.  The central
  1D velocity dispersion is consistent with a disc of moderate thickness,
  given that the central scale height is a few tens of parsecs, i.e.\ an
  order of magnitude smaller than the size of the disc (see Sect.\ 4.1).
\item The main molecular disc is also exponential, and has central surface
  density $\Sigma_{2}\simeq66\;\mbox{M}_{\odot}\,\mbox{pc}^{-2}$ and scale
  length $R_{2}\simeq2.9\,\mbox{kpc}$.  Note that $\Sigma_{2}$ is an
  extrapolated value, since the main and the nuclear discs join at a distance
  of about 1 kpc.  More precisely, the transition in surface density occurs
  at $R=R_{\mathrm{break}}\simeq0.75\,\mbox{kpc}$ and
  $\Sigma=\Sigma_{\mathrm{break}}\simeq50\;\mbox{M}_{\odot}\,\mbox{pc}^{-2}$.
  The 1D velocity dispersion is approximately constant up to the outermost
  radius of the analysis presented here (5 kpc), and has a value of about 10
  km\,s$^{-1}$.
\item These facts imply that the inner region of NGC 6946 is physically
  distinct from the rest of the disc, and that radial inflow and disc heating
  have both played a primary role in shaping the central kpc.  This is
  consistent with the fact that the dynamics of NGC 6946 is driven by three
  bars, and that the strong secondary bar extends up to $R\sim1\,\mbox{kpc}$.
  It is also consistent with the classical scenario of secular evolution in
  disc galaxies (see, e.g., Kormendy \& Kennicutt 2004; Kormendy 2013;
  Sellwood 2014).
\item The radial profile of the $\mathcal{Q}$ stability parameter is
  remarkably flat.  In spite of the fact that $\Sigma(R)$, $\kappa(R)$ and
  $\sigma(R)$ exhibit order-of-magnitude variations, $\mathcal{Q}(R)$
  fluctuates around a constant value, deviating from it by less than a factor
  of two!  This is an eloquent example of self-regulation: the disc of NGC
  6946 is driven by processes that keep it at a fairly constant stability
  level, except in the central kpc, where the value of $\mathcal{Q}$ is
  modulated by the nuclear starburst and inner bar within bar.  The fact that
  $\mathcal{Q}$ is well above unity, and on average 4--5, is far more
  intriguing.  We have discussed this issue in great detail and shown that
  the disc of NGC 6946 can indeed be close to marginal instability, but as a
  result of four destabilizing factors: stars, CO-dark molecular gas,
  anisotropy of the gas velocity dispersion and non-axisymmetric
  perturbations.
\item The radial profile of the characteristic instability wavelength,
  $\lambda_{\mathrm{c}}(R)$, is a powerful diagnostic that accurately
  predicts the sizes of starbursts and bars within bars.  Such a profile
  indicates that the nuclear starburst and the nuclear bar of NGC 6946 are
  intimately related, and were most likely generated by the same process of
  local gravitational instability in the disc, while the strong secondary bar
  has grown significantly after the onset of instability or was generated by
  larger-scale processes.
\end{itemize}

\section*{ACKNOWLEDGMENTS}

This work made use of data from the following surveys: BIMA SONG, `The BIMA
Survey of Nearby Galaxies' (Helfer et al.\ 2003); HERACLES, `The HERA CO-Line
Extragalactic Survey' (Leroy et al.\ 2009); and SINGS, `The \emph{SIRTF}
Nearby Galaxies Survey' (Kennicutt et al.\ 2003).  We are very grateful to
Oscar Agertz, Guillaume Drouart, Saladin Grebovi\'{c}, John Kormendy, Kirsten
Kraiberg Knudsen and Claudia Lagos for useful discussions.  We are also
grateful to an anonymous referee for constructive comments and suggestions,
and for encouraging future work on the topic.  KF acknowledges the
hospitality of the ESO Garching, where parts of this work were carried out.

\bsp

\label{lastpage}

\end{document}